\documentclass{article}
\usepackage[utf8]{inputenc}
\usepackage{authblk}
\usepackage{setspace}
\usepackage[margin=1.25in]{geometry}
\usepackage{graphicx}
\graphicspath{ {./figures/} }
\usepackage{subcaption}
\usepackage{amsmath}
\usepackage{hyperref}

\usepackage{mathrsfs}
\usepackage{amssymb}
\newtheorem{lemma}{Lemma}
\newtheorem{theorem}{Theorem}
\newtheorem{note}{Note}

\usepackage[style=nejm, 
citestyle=numeric-comp,
sorting=none]{biblatex}
\title{Event-Triggered Memory Control for Interval Type-2 Fuzzy Heterogeneous Multi-Agent Systems}

\author[1*$\dag$]{Sen Kong}

\date{}

\begin{document}

\maketitle

\begin{abstract}
This study explores the design of a memory-based dynamic event-triggered mechanisms (DETM) scheme for heterogeneous multi-agent systems (MASs) characterized by interval type-2 Takagi-Sugeno (IT2 T-S) fuzzy models. To address the complex nonlinear uncertainties inherent in such systems, discrete IT2 T-S fuzzy models are employed to accurately capture system dynamics. In response to the limitations on communication resources and computational capabilities within MASs, this research introduces a distributed DETM approach based on a dynamic threshold method. This mechanism effectively minimizes unnecessary communication while maintaining robust performance.
The proposed memory-based control strategy not only reduces the conservatism associated with controller design conditions but also enhances overall controller performance. Furthermore, leveraging a non-parallel distributed compensation (non-PDC) strategy, a novel derivation method is developed for controller design conditions that significantly decreases conservatism. This leads to sufficient conditions for the asymptotic stability of the closed-loop system. The designed distributed event-triggered controllers improve the overall performance of MASs, as evidenced by numerical simulations that validate the effectiveness of the proposed approach.
Overall, these findings advance the state-of-the-art in control strategies for heterogeneous MASs, offering practical solutions for real-world applications where resource constraints are critical.

\end{abstract}


\section{Introduction}

Multi-agent systems (MASs) have become a research hotspot due to their significant applications in distributed sensing, robotic cooperation, intelligent transportation, and other fields. However, controlling MASs typically faces two major challenges: first, heterogeneous MASs exhibit complex nonlinear uncertainties that make precise modeling difficult; second, the consistency of multi-agent collaborative control relies on data transmission between agents, but individual agents have limited communication, energy, and computational capabilities, which cannot support high-frequency data transmission and control output signal calculations. These issues represent bottlenecks in current MAS research.

To address the first challenge, fuzzy control systems provide an effective solution. Takagi-Sugeno (T-S) fuzzy control employs T-S fuzzy models to characterize nonlinear systems and designs corresponding controllers to ensure system stability. The T-S fuzzy model utilizes a set of If-Then rules with linear consequents, using membership functions to weight and normalize these sub-models, thereby describing nonlinear systems. This method allows for stability analysis and controller design using traditional linear control theory, leading to extensive research on stability analysis and controller synthesis for T-S fuzzy systems. Interval type-2 (IT2) T-S fuzzy models not only inherit the nonlinear processing capabilities of type-1 T-S fuzzy models but also enhance uncertainty representation by converting parameter uncertainties into membership function uncertainties, simplifying the calculation of type-2 fuzzy model membership functions and reducing computational complexity.

Controller design strategies can be categorized into Parallel Distributed Compensation (PDC) and Non-PDC strategies based on whether the membership functions and rule numbers of the fuzzy controller match those of the fuzzy system. The Non-PDC strategy allows designing independent membership functions for the fuzzy controller, enabling partial or complete mismatch of membership functions and varying rule numbers, thus providing greater design flexibility and robustness while potentially reducing controller design costs. However, this also introduces difficulties in applying methods that reduce design conservatism, as these methods are often tailored for PDC strategies.
Reference \cite{ref23} used a PDC strategy to design a type-1 fuzzy controller for truck-trailer systems. In contrast, \cite{ref24} introduced Non-PDC in interval type-2 fuzzy control, designing state feedback controllers with partially mismatched membership functions. \cite{ref25} extended this to fully mismatched membership functions, reducing design costs.


On the other hand, individual agents have limited communication, energy, and computational capabilities, which restrict their ability to support high-frequency data transmission and control output signal calculations. To mitigate this issue, time-triggered mechanisms (TTM) are employed to reduce communication frequency among agents \cite{ref61, ref62}. However, TTMs ignore real-time system conditions, potentially leading to redundant triggers.
In contrast, event-triggered mechanisms (ETM) compute trigger indicators based on system states and compare them with preset thresholds to determine if an event should be triggered. This approach more effectively conserves system resources, as demonstrated in \cite{ref63, ref64}. Moreover, \cite{ref65} proposed incorporating additional dynamic variables related to system states into the trigger threshold, thereby further reducing redundant event triggers.

To achieve fewer controller design conditions and improve robustness, scholars have proposed memory-based controllers that utilize historical state information to reduce conservatism. Studies show that memory-based controllers generally outperform non-memory controllers. For instance, \cite{ref35} designed a memory state feedback controller based on current and past system states; \cite{ref36, ref37} obtained periodic time-varying memory statistical feedback controllers and memory output feedback controllers through appropriate parameter selection; \cite{ref38} proposed a memory dynamic output feedback fuzzy controller, achieving good performance. However, these studies have not been sufficiently extended to IT2 T-S fuzzy systems with stronger nonlinear parameter uncertainty representation capabilities, leaving the design of memory output feedback controllers for such systems as an area for further exploration.

Some existing studies have explored dynamic ETM for cooperative control. For example, \cite{ref75} introduced a dynamic ETM for continuous type-1 T-S fuzzy MASs, while \cite{ref76} presented a discrete IT2 T-S fuzzy MASs method. However, both studies considered homogeneous agents and introduced some conservatism in deriving controller design conditions, limiting their application in heterogeneous MASs.

The main contributions of this study include:
\begin{enumerate}
    \item Designing a Non-PDC strategy-based IT2 T-S fuzzy heterogeneous MAS dynamic ETM with memory, addressing a less studied area.
    \item Resolving new state deviation variables and numerous coupling terms in controller stability analysis by employing appropriate mathematical tools and techniques, such as Linear Matrix Inequalities (LMIs), to reduce design conservatism.
    \item Handling membership functions containing agent index subscripts to obtain controller design conditions.
    \item Adopting a less conservative LMI transformation method to enhance controller design flexibility and effectiveness.
\end{enumerate}

The structure of this paper is organized as follows:
First, the paper presents the model representation for Discrete-Time IT2 T-S Fuzzy MASs. Based on this model, we establish a Distributed Dynamic Event-Triggered Mechanism with Memory (DETM). Subsequently, we design an IT2 T-S Fuzzy Controller with Memory. The proposed controller design and DETM are formulated into an augmented matrix form using the Kronecker product, leading to a novel closed-loop system representation.
Next, under this new closed-loop system framework, we address the challenges in controller stability analysis by handling new state deviation variables and numerous coupling terms. We also appropriately manage the agent index subscripts within the membership functions. To derive the controller design conditions, we adopt a less conservative LMI transformation method, which enhances the flexibility and effectiveness of the design process.
Finally, the effectiveness of the proposed approach is demonstrated through numerical simulations involving heterogeneous MASs.

\subsection*{Notations:}

Let $\mathbb{R}^{n}$ denote the $n$-dimensional Euclidean space. The transpose of a matrix $Q$ is denoted by $Q^{\top}$. For a matrix $Q$, $Sym\{Q\}$ represents the symmetric part of $Q$, defined as $Q + Q^{\top}$. 
For a matrix $Q$, the notations $Q > 0$ and $Q \geq 0$ indicate that $Q$ is positive definite and positive semi-definite, respectively. Similarly, $Q < 0$ and $Q \leq 0$ denote that $Q$ is negative definite and negative semi-definite, respectively.
Given matrices $Q_1, Q_2, \ldots, Q_n$, $diag\{Q_1, Q_2, \ldots, Q_n\}$ denotes a block-diagonal matrix with the block-diagonal elements being $Q_1, Q_2, \ldots, Q_n$.
In symmetric matrices, the symbol $\star$ indicates the terms that can be inferred by symmetry.

\section{Model Description and Problem Formulation}

This section begins with an introduction to discrete-time interval type-2 T-S fuzzy multi-agent systems, followed by a presentation of the distributed dynamic event-triggered mechanism. Subsequently, a memory-based fuzzy controller is designed. Finally, the expression for the closed-loop system is provided.

\subsection*{Discrete-Time Interval Type-2 T-S Fuzzy Multi-Agent Systems:}

Consider the discrete-time IT2 T-S fuzzy MASs with $N$ identical agents, and the mathematical model of the $i$th agent can be  described by the following IF-THEN rules.

$$
\textbf{Plant Rule } \mathscr{P}^{i}_{l_i}\textbf{: }\textbf{IF } \chi_{1}^{i}(x(t)) \textbf{ is } {W}_{1}^{l_i}\textbf{ and } \chi_{2}^{i}(x(t)) \textbf{ is }{W}_{2}^{l_i} \textbf{ and } \cdots \textbf{ and }
$$
$$
\chi_{\rho}^{i}(x(t))  \textbf{ is } {W}_{\rho}^{l_i}, \textbf{ THEN } \nonumber \ \ \ \ \ \ \ \ \ \ \ \ \ \
$$
\begin{equation}
x_i(t+1)= A_{l_i}^i x_i(t) + B_{l_i}^i u_i(t)
\end{equation}

where
$\mathscr{P}^{i}_{l_i}$ symbolizes the $l_i$th fuzzy inference rule of the $i$th agent, and $p_i$ denotes the number of IF-THEN rules of the $i$th agent;
$\chi^{i}(x(t))=\left[\chi_{1}^{i}(x(t)), \chi_{2}^{i}(x(t)), \ldots, \chi_{\omega}^{i}(x(t))\right]$ are measurable premise variables of the system;
$W_{\varrho}^{i}(\varrho=1,2, \ldots, \rho)$ are type-2 fuzzy sets;
$x_i(t) \in \Re^{n_{x}}$ and $u_i(t) \in \Re^{n_{u}}$ represent the system state and the control input of the $i$th agent;
$A_{l_i}^i$ and $B_{l_i}^i$ are known real constant matrices, which are used to represent the $l_i$th local model of the $i$th agent.

The weight of the $l_i$th rule is belonging to the interval sets:
\begin{align}
    \mathscr{W}_{l_i}=\left[
    \begin{array}{c c}
        \underline{w}_{l_i}(x(t)) & \bar{w}_{l_i}(x(t))
    \end{array}\right], \ l_i \in \mathscr{P}_i
\end{align}
with
\begin{equation}
    \begin{gathered}
        \underline{w}_{l_i}(x(t))=\prod_{\varrho=1}^{\rho} \underline{\varpi}_{W_{\varrho}^{l_i}(\chi^{i}(x(t)))} \geq 0,\ 
        \bar{w}_{l_i}(x(t))=\prod_{\varrho=1}^{\rho} \bar{\varpi}_{W_{\varrho}^{l_i}(\chi^{i}(x(t)))} \geq 0, \\
        \bar{\varpi}_{W_{\varrho}^{l_i}(\chi^{i}(x(t)))} \geq \underline{\varpi}_{W_{\varrho}^{l_i}(\chi^{i}(x(t)))} \geq 0, \ 
        \bar{w}_{l_i}(x(t)) \geq \underline{w}_{l_i}(x(t))  \geq 0
    \end{gathered}
\end{equation}

where $\underline{w}_{l_i}(x(t))$ and $\bar{w}_{l_i}(x(t))$ are the lower and upper membership functions, respectively;
$\underline{\varpi}_{W_{\varrho}^{l_i}(\chi^{i}(x(t)))}$ and $\bar{\varpi}_{W_{\varrho}^{l_i}(\chi^{i}(x(t)))}$ being the lower and upper grades of the membership of $\chi_{\varrho}^{i}(x(t))$
in $M_{\varrho}^{l_i}$, respectively.

Let $w_{l_i}(x(t))$ represent the normalized membership function satisfying
\begin{equation}
    \begin{gathered}
        w_{l_i}(x(t)) = \frac{\underline{\varsigma}_{l_i}(x(t)) \underline{w}_{l_i}(x(t)) + \bar{\varsigma}_{l_i}(x(t)) \bar{w}_{l_i}(x(t))}{\sum_{k \in \mathscr{P}_i}\left(\underline{\varsigma}_{k}(x(t)) \underline{w}_{k}(x(t))+\bar{\varsigma}_{k}(x(t)) \bar{w}_{k}(x(t))\right)},\\
        w_{l_i}(x(t)) \geq 0,\ \sum_{l_i \in \mathscr{P}_i} w_{l_i}(x(t))=1
    \end{gathered}
\end{equation}
with $\underline{\varsigma}_{l_i}(x(t))$, $\bar{\varsigma}_{l_i}(x(t))$ are satisfying
\begin{equation}
    0 \leq \underline{\varsigma}_{l_i}(x(t)), \bar{\varsigma}_{l_i}(x(t)) \leq 1,\ 
    \underline{\varsigma}_{l_i}(x(t))+\bar{\varsigma}_{l_i}(x(t))=1.
\end{equation}

It is worth mentioning that the nonlinear functions $\underline{\varsigma}_{l_i}(x(t))$ and $\bar{\varsigma}_{l_i}(x(t))$ describe the parameter uncertainties, and they are unessential to be known.

The global fuzzy model can be inferred as follows:
\begin{equation}
    x_i(t+1)= \sum_{l_i=1}^{p_i} m^{i}_{l_i} \left( A_{l_i}^i x_i(t) +B_{l_i}^i u_i(t) \right)
\end{equation}

\subsection*{Distributed Dynamic Event-Triggered Mechanism with Memory:}

To reduce communication overhead, a direct and natural approach is to decrease the number of data release triggering events. To alleviate communication overhead and reduce the update frequency of controllers, this chapter establishes a DETM based on dynamic threshold strategies.

Similar to the method in article \cite{ref76}, the triggering time sequence is defined as $\{t_k^i, k=1,2,\cdots\}$. The state signal $x_i(t)$ at the current moment and the state signal $x_i(t_k^i)$ at the previous triggering moment are expressed as the difference $e_i(t) = x_i(t_k^i) - x_i(t)$.

For the $i$th intelligent body, if the previous triggering moment is $t_k^i$, then the next triggering moment $t_{k+1}^i$ can be determined by the following DETM condition:

\begin{equation}
\label{eq:3-7}
t_{k+1}^i = \min \left\{ t \mid t > t_k^i, e_i^T(t)\Omega_i e_i(t) \geq \Delta_i \delta_i^T(t_k^i)\Omega_i \delta_i(t_k^i) \right\},
\end{equation}

where $\Omega_i$ represents the weight matrix of the event-triggering mechanism, the dynamic threshold variable of the event-triggering mechanism, $\Delta_i$, depends on $e_i^T(t)e_i(t)$ and is given by
$\Delta_i = \alpha_i \left(1 - \beta_i \tanh(e_i^T(t)e_i(t) - \theta_i)\right)$,
where $\alpha_i$, $\beta_i$, and $\theta_i$ are the given threshold parameters.

Additionally, the variable $\delta_i(t_k^i)$ can be calculated using the following formula:

\begin{equation}
\delta_i(t_k^i) = \sum_{j=1}^{N} a_{ij}(x_i(t_k^i) - x_j(t_k^i)) + b_i x_i(t_k^i),
\end{equation}

where $a_{ij}$ and $b_i$ represent the weights of the formation maintenance behavior and state maintenance behavior in the variable $\delta_i(t_k^i)$.

Considering $x_i(t_k^i) = e_i(t) + x_i(t)$, the above equation can be rewritten as:

\begin{equation}
\delta_i(t_k^i) = \sum_{j=1}^{N} a_{ij}((e_i(t) + x_i(t)) - (e_j(t) + x_j(t))) + b_i(e_i(t) + x_i(t)).
\end{equation}

In the subsequent derivation process, a useful and important condition is the event-triggering condition designed in the expected DETM. The specific description of the event-triggering condition for the $i$th intelligent body is as follows:

\begin{equation}
e_i^T(t)\Omega_i e_i(t) \leq \Delta_i \delta_i^T(t_k^i)\Omega_i \delta_i(t_k^i).
\end{equation}

\begin{note}\label{note:1}
    To better adapt to practical situations, this chapter designs a more practical distributed dynamic event-triggering mechanism for discrete-time multi-agent systems. The mechanism employs a dynamic threshold variable given by:
    $
    \Delta_i = \alpha_i \left(1 - \beta_i \tanh(e_i^T(t)e_i(t) - \theta_i)\right).
    $
    Clearly, the larger the basic threshold parameter $\alpha_i$, the larger $\Delta_i$ becomes, making the mechanism less likely to trigger. The threshold parameter $\beta_i$ is used to further regulate $\Delta_i$, making its changes smoother. The variation of the dynamic threshold $\Delta_i$ mainly depends on the measurement error variable $e_i(t)$, which changes with the degree of system state variations.
    Based on the properties of the $\tanh(\bullet)$ function, it can be observed that when $e_i^T(t)e_i(t) > \theta_i$, the system's state fluctuates significantly, and the actual dynamic threshold $\Delta_i$ decreases, allowing more frequent transmission of system information. This means that stabilizing the system state will be more valuable. Conversely, when $e_i^T(t)e_i(t) < \theta_i$, the system's state is relatively stable, and the actual dynamic threshold $\Delta_i$ increases, reducing the number of triggers. This means that more resources will be conserved.
    Specifically, if $\beta_i = 0$, the dynamic event-triggering mechanism degenerates into a static event-triggering mechanism; if $\alpha_i = 0$, the dynamic event-triggering mechanism fails, and events are triggered at every time instant.
\end{note}

\subsection*{Design of Interval Type-2 T-S Fuzzy Controller with Memory:}

Based on the discussion above, for a multi-agent system consisting of $N$ heterogeneous intelligent agents, a $\kappa$-order interval type-2 T-S fuzzy controller with memory state feedback can be described as follows.

Using an interval type-2 T-S fuzzy system to describe it, the mathematical model of the $i$-th intelligent agent can be expressed by the following If-Then rules:

Rule $C_{s_i}^i$: If $\tau_1^i(x(t))$ is $M_1^{s_i}$, and $\tau_2^i(x(t))$ is $M_2^{s_i}$, and $\cdots$, and $\tau_\psi^i(x(t))$ is $M_\psi^{s_i}$, then

\begin{equation}
u_i(t) = \sum_{h=1}^\kappa K_{s_i}^{i(h)} \delta_i(t_k^i - h + 1)
\end{equation}

where $C_{s_i}^i$ represents the $s_i$-th sub-fuzzy controller of the $i$-th intelligent agent, $s_i \in C_i = \{1, 2, \cdots, q_i\}$, and $q_i$ is the number of fuzzy inference rules of the $i$-th intelligent agent; $\tau(x(t)) = [\tau_1(x(t)), \tau_2(x(t)), \cdots, \tau_\psi(x(t))]$ is the measurable premise variable of the system; $M_\phi^{s_i} (\phi = 1, 2, \cdots, \psi)$ is the interval type-2 fuzzy set; $K_{s_i}^{i(h)}$ is the gain of the designed controller, representing the sub-controller dependent on the $h$-th historical time point of the $s_i$-th local controller of the $i$-th intelligent agent.

Next, construct the global state feedback controller as follows:

\begin{equation}
u_i(t) = \sum_{s_i=1}^{q_i} m_{s_i}^i \sum_{h=1}^\kappa K_{s_i}^{i(h)} \delta_i(t_k^i - h + 1)
\end{equation}

where $m_{s_i}^i$ represents the membership degree of the interval type-2 fuzzy function.

To better express the characteristics of the controller considering historical state information, introduce the following augmented variables:

$$
\begin{cases}
\tilde{x}_i(t) = [x_i^T(t-\kappa+1) \cdots x_i^T(t-h+1) \cdots x_i^T(t-1) \ x_i^T(t)]^T, \\
\tilde{e}_i(t) = [e_i^T(t-\kappa+1) \cdots e_i^T(t-h+1) \cdots e_i^T(t-1) \ e_i^T(t)]^T, \\
\tilde{\delta}_i(t_k^i) = [\delta_i^T(t_k^i-\kappa+1) \cdots \delta_i^T(t_k^i-h+1) \cdots \delta_i^T(t_k^i-1) \ \delta_i^T(t_k^i)]^T.
\end{cases}
$$

The relationship between variables can be rewritten in the following form:

\begin{equation}
\tilde{\delta}_i(t_k^i) = \sum_{j=1}^N a_{ij} ((\tilde{e}_i(t) + \tilde{x}_i(t)) - (\tilde{e}_j(t) + \tilde{x}_j(t))) + b_i (\tilde{e}_i(t) + \tilde{x}_i(t)).
\end{equation}

Thus, similarly, the global state feedback controller can be rewritten in the following form:

\begin{equation}
u_i(t) = \sum_{s_i=1}^{q_i} n_{s_i}^i \tilde{K}_{s_i}^i \tilde{\delta}_i(t_k^i)
\end{equation}

where $\tilde{K}_{s_i}^i = [K_{s_i}^{i(\kappa)} \cdots K_{s_i}^{i(h)} \cdots K_{s_i}^{i(2)} \ K_{s_i}^{i(1)}]$.

\begin{note}
    \label{note:2} 
    In this chapter, $\kappa$ represents the upper limit of historical time data. If $\kappa = 1$, the memory controller will degenerate into a traditional non-memory controller. More specifically, when $t-\kappa+1 < 0$, assume $x(t-\kappa+1) = x(t_0)$, where $x(t_0)$ represents the initial state of the system.
\end{note}

After the above discussion, the closed-loop system of this model can be described as follows:

\begin{equation}
\label{eq:3-16}
x_i(t+1) = \sum_{l_i=1}^{p_i} w_{l_i}^i \left( A_{l_i}^i x_i(t) + B_{l_i}^i \sum_{s_i=1}^{q_i} m_{s_i}^i \tilde{K}_{s_i}^i \tilde{\delta}_i(t_k^i) \right)
\end{equation}



Considering the augmented variables $\tilde{x}_i(t)$ described above, the closed-loop system (\ref{eq:3-16}) can be rewritten as follows:

\begin{equation}
\tilde{x}_i(t+1) = \sum_{l_i=1}^{p_i} w_{l_i}^i \left( \tilde{A}_{l_i}^i \tilde{x}_i(t) + \tilde{B}_{l_i}^i \sum_{s_i=1}^{q_i} m_{s_i}^i \tilde{K}_{s_i}^i \tilde{\delta}_i(t_k^i) \right)
\label{eq:3-17}
\end{equation}

where

\begin{equation}
\tilde{A}_{l_i}^i = \begin{bmatrix}
0 & I \\
0 & [0 \ A_{l_i}^i]
\end{bmatrix}, \quad
\tilde{B}_{l_i}^i = \begin{bmatrix}
0 \\
B_{l_i}^i
\end{bmatrix}.
\end{equation}

Therefore, considering all intelligent agents under the multi-agent system and based on the properties of the Kronecker product, the closed-loop system (\ref{eq:3-17}) can be rewritten as follows:

\begin{equation}
\label{eq:3-19}
\mathbf{x}(t+1) = \sum_{i=1}^{N} \left( \sum_{l_i=1}^{p_i} m_{l_i}^i \left( \bar{\mathbb{A}}_{l_i}^i \mathbf{x}(t) + \sum_{s_i=1}^{q_i} n_{s_i}^i \bar{\mathbb{B}}_{l_i}^i (L_i + R_i) \otimes \tilde{K}_{s_i}^i (\mathbf{x}(t) + \mathbf{e}(t)) \right) \right)
\end{equation}

where


\begin{equation}
\begin{cases}
\bar{\mathbb{A}}_{l_i}^i = \text{diag} \left\{ \underbrace{0 \cdots 0}_{i-1}, \tilde{A}_{l_i}^i, \underbrace{0 \cdots 0}_{N-i} \right\}, \\
\bar{\mathbb{B}}_{l_i}^i = \text{diag} \left\{ \underbrace{0 \cdots 0}_{i-1}, \tilde{B}_{l_i}^i, \underbrace{0 \cdots 0}_{N-i} \right\}, \\
L_i = \text{diag} \left\{ \underbrace{0 \cdots 0}_{i-1}, \bar{a}_i, \underbrace{0 \cdots 0}_{N-i} \right\} - \begin{bmatrix}
0 & \cdots & 0 \\
\vdots & \ddots & \vdots \\
a_{i1} & \cdots & a_{iN} \\
\vdots & \ddots & \vdots \\
0 & \cdots & 0
\end{bmatrix}, \\
R_i = \text{diag} \left\{ \underbrace{0 \cdots 0}_{i-1}, b_i, \underbrace{0 \cdots 0}_{N-i} \right\}
\end{cases}
\end{equation}

and $\bar{a}_i = \sum_{j=1}^{N} a_{ij}$.

Considering $\sum_{l_i=1}^{p_i} m_{l_i}^i = \sum_{s_i=1}^{q_i} n_{s_i}^i = 1$, $\forall i = 1, \ldots, N$, the closed-loop system (\ref{eq:3-19}) can be rewritten as follows to better handle the membership functions of the system and controller:

\begin{equation}
\mathbf{x}(t+1) = \prod_{i'=1}^{N} \left( \sum_{l_{i'}=1}^{p_{i'}} m_{l_{i'}}^{i'} \right) \prod_{i''=1}^{N} \left( \sum_{s_{i''}=1}^{q_{i''}} n_{s_{i''}}^{i''} \right) \sum_{i=1}^{N} \left( \bar{\mathbb{A}}_{l_i}^i \mathbf{x}(t) + \bar{\mathbb{B}}_{l_i}^i (L_i + R_i) \otimes \tilde{K}_{s_i}^i (\mathbf{x}(t) + \mathbf{e}(t)) \right).
\label{eq:3-22}
\end{equation}

For the derivation of subsequent theoretical results, this chapter provides the following important lemma:

\begin{lemma}
    \label{lemma:1}
    \cite{ref77}
    For any matrices $\mathbb{L}_{l_i s_i}^i$, $\Psi$, $\mathbb{R}_{l_i s_i}^i > 0$, and constants $m_{l_{i'}}^{i'}, n_{s_{i''}}^{i''} \in [0, 1]$, the following matrix inequality holds:
    \begin{equation}
        \begin{split}
            & \left[ \prod_{i'=1}^{N} \left( \sum_{l_{i'}=1}^{p_{i'}} m_{l_{i'}}^{i'} \right) \prod_{i''=1}^{N} \left( \sum_{s_{i''}=1}^{q_{i''}} n_{s_{i''}}^{i''} \right) \mathbb{L}_{l_i s_i}^i \right]^T \Psi \times \left[ \prod_{i'=1}^{N} \left( \sum_{l_{i'}=1}^{p_{i'}} m_{l_{i'}}^{i'} \right) \prod_{i''=1}^{N} \left( \sum_{s_{i''}=1}^{q_{i''}} n_{s_{i''}}^{i''} \right) \mathbb{R}_{l_i s_i}^i \right] \\
            & \leq \frac{1}{2} \prod_{i'=1}^{N} \left( \sum_{l_{i'}=1}^{p_{i'}} m_{l_{i'}}^{i'} \right) \prod_{i''=1}^{N} \left( \sum_{s_{i''}=1}^{q_{i''}} n_{s_{i''}}^{i''} \right) \left[ \mathbb{L}_{l_i s_i}^{i^T} \Psi \mathbb{L}_{l_i s_i}^i + \mathbb{R}_{l_i s_i}^{i^T} \Psi \mathbb{R}_{l_i s_i}^i \right].
        \end{split}
    \end{equation}
\end{lemma}

\subsection*{Problem Description:}

The objective of this chapter is to synthesize a memory state-feedback controller that ensures the closed-loop system (\ref{eq:3-22}) is globally asymptotically stable.

\section{Main Results}

\subsubsection*{Event-Triggered Memory Controller Analysis and Synthesis:}

\begin{theorem}
    \label{theorem:1} 
    Given parameters $a_{ij}$, $b_i$, $\sigma$, $\alpha_i$, $\beta_i$, and a given matrix $H$, the closed-loop system is asymptotically stable if there exist symmetric matrix variables $P$, $\Omega_i$, and $X_s^i$, and matrix variables $\tilde{Y}_s^i$ that satisfy the following linear matrix inequality:
    \begin{equation}
    P > 0,
    \end{equation}
    
    \begin{equation}
    \label{eq:3-25}
    {\Xi _{{l_i}{s_i}}} = 
    \begin{bmatrix}
    -P + {\Lambda _E} & \star & \star & \star \\
    {\Lambda _E} & {\Lambda _E} - {\Upsilon _\Omega} & \star & \star \\
    {\Xi _{{l_i}{s_i}}^{(5)}} & {\Xi _{{l_i}{s_i}}^{(1)}} & -P & \star \\
    {\sigma \Xi _{{s_i}}^{(4)}} & {\sigma \Xi _{{s_i}}^{(4)}} & {\Xi _{{l_i}{s_i}}^{(2)}} & -\sigma \Xi _{{s_i}}^{(3)} - \sigma (\Xi _{{s_i}}^{(3)})^T
    \end{bmatrix}
    < 0,
    \end{equation}
    
    where
    \begin{equation}
    \left\{
    \begin{aligned}
    &{\Lambda _E} = (L + R)^T \Gamma_\Omega (L + R), \\
    &{\Upsilon _\Omega} = \text{diag} \{ H^T \Omega_1 H, H^T \Omega_2 H, \cdots, H^T \Omega_N H \}, \\
    &{\Xi _{{l_i}{s_i}}^{(1)}} = \sum_{i=1}^N \left( \mathbb{B}_l^i (L_i + R_i) \otimes \tilde{Y}_{s_i}^i \right), \\
    &{\Xi _{{l_i}{s_i}}^{(2)}} = 
    \begin{bmatrix}
    \left( P \mathbb{B}_l^1 (L_1 + R_1) \otimes I - \mathbb{B}_l^1 (L_1 + R_1) \otimes X_{s_1}^1 \right)^T \\
    \vdots \\
    \left( P \mathbb{B}_l^N (L_N + R_N) \otimes I - \mathbb{B}_l^N (L_N + R_N) \otimes X_{s_N}^N \right)^T
    \end{bmatrix}, \\
    &{\Xi _{{s_i}}^{(3)}} = \text{diag} \left\{ I \otimes X_{s_1}^i, \cdots, I \otimes X_{s_N}^i \right\}, \\
    &{\Xi _{{s_i}}^{(4)}} = 
    \begin{bmatrix}
    I \otimes \tilde{Y}_{s_1}^1 \\
    \vdots \\
    I \otimes \tilde{Y}_{s_N}^N
    \end{bmatrix}, \\
    &{\Xi _{{l_i}{s_i}}^{(5)}} = \sum_{i=1}^N \left( \bar{A}_l^i + \mathbb{B}_l^i (L_i + R_i) \otimes \tilde{Y}_{s_i}^i \right), \\
    &L = \sum_{i=1}^N L_i, \quad R = \sum_{i=1}^N R_i, \\
    &\Gamma_\Omega = \text{diag} \{ \alpha_1 (1 + \beta_1) H^T \Omega_1 H, \alpha_2 (1 + \beta_2) H^T \Omega_2 H, \cdots, \alpha_N (1 + \beta_N) H^T \Omega_N H \}.
    \end{aligned}
    \right.
    \label{eq:3-26}
    \end{equation}
    
    Additionally, the controller gain can be computed by $\tilde{K}_{s_i}^i = X_{s_i}^{i^{-1}} \tilde{Y}_{s_i}^i$.
\end{theorem}

\textbf{Proof:} Now, construct the Lyapunov function as follows:
\begin{equation}
V(t) = \mathbf{x}^T(t) P \mathbf{x}(t),
\end{equation}
where $ P \in \Re^{N_{kn_x} \times N_{kn_x}} $.

Then the difference of the Lyapunov function is
\begin{equation}\label{eq:3-28}
\Delta V(t) = \mathbf{x}^T(t+1) P \mathbf{x}(t+1) - \mathbf{x}^T(t) P \mathbf{x}(t).
\end{equation}

On the other hand, consider the dynamic event-triggering condition used in this chapter $ e_i^T(t) \Omega_i e_i(t) \leq \Delta_i \delta_i^T(t_k^i) \Omega_i \delta_i(t_k^i) $. By considering the augmented variables at historical times, this condition can be rewritten as:
\begin{equation}
\label{eq:3-29}
\tilde{e}_i^T(t) H^T \Omega_i H \tilde{e}_i(t) \leq \Delta_i \tilde{\delta}_i^T(t_k^i) H^T \Omega_i H \tilde{\delta}_i(t_k^i).
\end{equation}

\textbf{Note 3:} In equation (\ref{eq:3-29}), the use of matrix $ H $ essentially aims to augment matrix $ \Omega_i $. When matrix $ H $ is set to $H=\left[ \begin{matrix}
   \underbrace{\begin{matrix}
   0 & \cdots  & 0  \\
\end{matrix}}_{\kappa -1} & I  \\
\end{matrix} \right]$, it corresponds to the memoryless distributed dynamic event-triggering mechanism method; when matrix $ H $ is set to $H=\left[ \underbrace{\begin{matrix}
   {{{\hat{H}}}_{\kappa }} & \cdots  & {{{\hat{H}}}_{1}}  \\
\end{matrix}}_{\kappa } \right]$, it corresponds to the memory-based distributed dynamic event-triggering mechanism method in this chapter. Since the latest information is considered more valuable, so $ \hat{H}_{h+1} \leq \hat{H}_h \leq \hat{H}_{h-1}, \forall h = 1, 2, \ldots, \kappa $, then the event-triggering mechanism (\ref{eq:3-7}) should also be modified to a memory-based event-triggering mechanism, i.e., introducing the augmented variables including matrix $ H $, or expanding the dimension of variable matrix $ \Omega_i $ by a factor of $ \kappa $.

In the memory-based dynamic event-triggering condition $ e_i^T(t) \Omega_i e_i(t) \leq \Delta_i \delta_i^T(t_k^i) \Omega_i \delta_i(t_k^i) $, there is an implicit condition, i.e., $\Delta_i = \alpha_i (1 - \beta_i \tanh(e_i^T(t) e_i(t) - \theta_i)) \leq \alpha_i (1 + \beta_i)$. In the subsequent part, use $\alpha_i (1 + \beta_i)$ to replace the nonlinear function $\alpha_i (1 - \beta_i \tanh(e_i^T(t) e_i(t) - \theta_i))$ for derivation.

Next, consider the entire multi-agent system, and the following judgment condition can be obtained:
\begin{equation}\label{eq:3-30}
\mathbf{e}^T(t) Y_\Omega \mathbf{e}(t) \leq (\mathbf{x}(t) + \mathbf{e}(t))^T ((L + R)^T \Gamma_\Omega (L + R)) (\mathbf{x}(t) + \mathbf{e}(t))
\end{equation}
where $L$, $R$, $\Gamma_\Omega$, and $Y_\Omega$ are defined in equation (\ref{eq:3-26}).

Next, let $\varphi(t) = [\mathbf{x}^T(t), \mathbf{e}^T(t)]^T$, and combining equations (\ref{eq:3-28}) and (\ref{eq:3-30}) yields the following condition:
\begin{equation}
\begin{split}
\Delta V(t) &\leq \mathbf{x}^T(t+1) P \mathbf{x}(t+1) - \mathbf{x}^T(t) P \mathbf{x}(t) \\
&+ (\mathbf{x}(t) + \mathbf{e}(t))^T ((L + R)^T \Gamma_\Omega (L + R)) (\mathbf{x}(t) + \mathbf{e}(t)) - \mathbf{e}_i^T(t) Y_\Omega \mathbf{e}_i(t) \\
&\leq \prod_{l=1}^{N} \left( \sum_{l'_l=1}^{p_l} m_{l'_l}^l \right) \prod_{r'=1}^{N} \left( \sum_{s'_{r'}=1}^{q_{r'}} n_{s'_{r'}}^{r'} \right) \varphi^T(t) \Theta_{l_s} \varphi(t)
\end{split}
\end{equation}
where the subscript of matrix $\Theta_{l_s}$ is actually $l_1, l_2, \cdots, l_N, s_1, s_2, \cdots, s_N$, and $l_s$ is used as a shorthand. Additionally,
\begin{equation}
    \label{eq:3-32}
    \Theta_{l_s} = 
    \begin{bmatrix}
        -P + \Lambda_{A_{l|s}}^T P \Lambda_{A_{l|s}} + \Lambda_E & * \\
        \Lambda_{B_{l|s}}^T P \Lambda_{A_{l|s}} + \Lambda_E & \Lambda_{B_{l|s}}^T P \Lambda_{B_{l|s}} + \Lambda_E - Y_\Omega
    \end{bmatrix},
\end{equation}
where the variable $\Lambda_E$ is already defined in equation (\ref{eq:3-26}), and additionally,
\begin{equation}
\begin{cases}
\Lambda_{A_{l|s}} = \sum_{i=1}^{N} \left( \mathbb{A}_{l_i}^i + \mathbb{B}_{l_i}^i (L_i + R_i) \otimes \tilde{K}_{s_i}^i \right), \\
\Lambda_{B_{l|s}} = \sum_{i=1}^{N} \left( \mathbb{B}_{l_i}^i (L_i + R_i) \otimes \tilde{K}_{s_i}^i \right).
\end{cases}
\end{equation}

By applying the Schur complement theorem to equation (\ref{eq:3-32}), we can obtain the following:
\begin{equation}
\begin{bmatrix}
-P + \Lambda_E & \star & \star \\
\Lambda_E & \Lambda_E - Y_\Omega & \star \\
\Lambda_{A_{l|s_i}} & \Lambda_{B_{l|s_i}} & -P^{-1}
\end{bmatrix}.
\end{equation}

Multiplying the above equation on the left and right by $\text{diag}\{I, I, P\}$ and its transpose, we can obtain the following:
\begin{equation}
    \label{eq:3-35}
    \begin{bmatrix}
        -P + \Lambda_E & \star & \star \\
        \Lambda_E & \Lambda_E - Y_\Omega & \star \\
        P \Lambda_{A_{l|s_i}} & P \Lambda_{B_{l|s_i}} & -P
    \end{bmatrix}.
\end{equation}

To ensure that $\mathbf{\Theta}_{l_i s_i} < 0$, we revisit equation (\ref{eq:3-35}) and analyze $\Xi_{l_i s_i}$. Considering $\tilde{K}_{s_i}^i = X_{s_i}^{i-1} \tilde{Y}_{s_i}^i$, the coupled variables $P \Lambda_{B_{l_i s_i}}$ and $P \Lambda_{A_{l_i s_i}}$ in equation (\ref{eq:3-35}) can be transformed as follows:

\begin{equation}
    \label{eq:3-36}
    \begin{aligned}
        P \Lambda_{B_{l_i s_i}} &= \sum_{i=1}^N \left( P \mathbb{B}_l^i (L_i + R_i) \otimes \tilde{K}_{s_i}^i \right) \\
        &= \sum_{i=1}^N \left( \mathbb{B}_l^i (L_i + R_i) \otimes \tilde{Y}_{s_i}^i + \left( P \mathbb{B}_l^i (L_i + R_i) \otimes I - \mathbb{B}_l^i (L_i + R_i) \otimes X_{s_i}^i \right) \right) \\
        &= \left( I \otimes X_{s_i}^{i-1} \right) \left( I \otimes \tilde{Y}_{s_i}^i \right) \\
        &= \Xi_{l_i s_i}^{(1)} + \Xi_{l_i s_i}^{(2)T} \Xi_{s_i}^{(3)-1} \Xi_{s_i}^{(4)}, \\
        P \Lambda_{A_{l_i s_i}} &= \sum_{i=1}^N \left( P \mathbb{A}_l^i + P \mathbb{B}_l^i (L_i + R_i) \otimes \tilde{K}_{s_i}^i \right) \\
        &= \sum_{i=1}^N \left( P \mathbb{A}_l^i + \mathbb{B}_l^i (L_i + R_i) \otimes \tilde{Y}_{s_i}^i + \left( P \mathbb{B}_l^i (L_i + R_i) \otimes I - \mathbb{B}_l^i (L_i + R_i) \otimes X_{s_i}^i \right) \right) \\
        &= \left( I \otimes X_{s_i}^{i-1} \right) \left( I \otimes \tilde{Y}_{s_i}^i \right) \\
        &= \Xi_{l_i s_i}^{(5)} + \Xi_{l_i s_i}^{(2)T} \Xi_{s_i}^{(3)-1} \Xi_{s_i}^{(4)},
    \end{aligned}
\end{equation}

where the variables $\Xi_{l_i s_i}^{(1)}$, $\Xi_{l_i s_i}^{(2)}$, $\Xi_{s_i}^{(3)}$, $\Xi_{s_i}^{(4)}$, and $\Xi_{l_i s_i}^{(5)}$ are defined in equation (\ref{eq:3-26}).

By applying equation (\ref{eq:3-36}) and Theorem~\ref{theorem:1} to equation (\ref{eq:3-35}), if there exists $\sigma > 0$, we will obtain the conclusion: condition (\ref{eq:3-25}) can ensure that $\Theta_{l_i s_i} < 0$, i.e., the following holds:
\begin{equation}
\Delta V(t) \leq \prod_{i'=1}^{N} \left( \sum_{l'_i=1}^{p_{i'}} m_{l'_i}^{i'} \right) \prod_{i''=1}^{N} \left( \sum_{s'_{i''}=1}^{q_{i''}} n_{s'_{i''}}^{i''} \right) \varphi^T(t) \Theta_{l_i s_i} \varphi(t) < 0.
\tag{3-37}
\end{equation}

Summarizing the above derivation, the theorem ensures that $\Delta V(t) < 0$ holds, which means the interval type-2 T-S fuzzy MAS (\ref{eq:3-22}) is asymptotically stable. Proof completed.

\begin{figure}[!ht]
    \centering
    \includegraphics[width=0.6\textwidth]{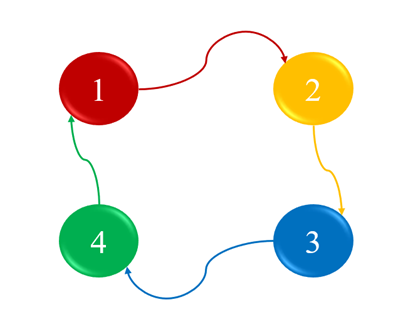}
    \caption{The communication topology of the multi-agent system.}
    \label{fig:4.1}
\end{figure}

\section{Simulation Studies}

This section verifies the effectiveness and superiority of the distributed dynamic event-triggering mechanism and memory-based control strategy through an example of an interval type-2 T-S fuzzy heterogeneous multi-agent system.

\begin{figure}[!ht]
    \centering
    \includegraphics[width=0.8\textwidth]{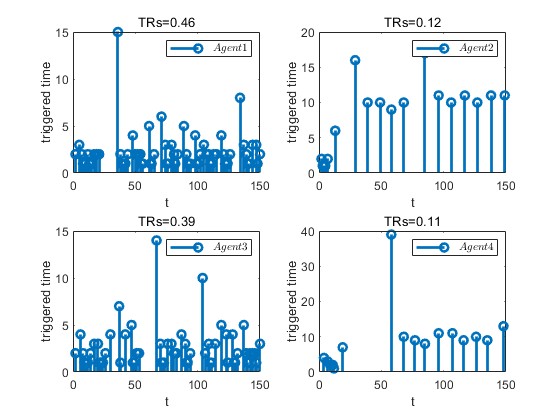}
    \caption{The event triggered conditions of the multi-agent system.}
    \label{fig:4.2}
\end{figure}

A communication topology with $ N = 4 $ is shown in Figure \ref{fig:4.1} for an interval type-2 fuzzy heterogeneous multi-agent system. The Laplacian matrix is given as follows:
$$
L = \begin{bmatrix}
1 & 0 & 0 & -1 \\
-1 & 1 & 0 & 0 \\
0 & -1 & 1 & 0 \\
0 & 0 & -1 & 1
\end{bmatrix},
$$
and $ b_1 = b_2 = b_3 = b_4 = 1 $. The system matrices for the first agent are:
$$
A_1^1 = \begin{bmatrix}
0.8 & 0.01 \\
0.1 & 1
\end{bmatrix}, \quad
A_2^1 = \begin{bmatrix}
0.85 & 0.01 \\
0.1 & 1
\end{bmatrix}, \quad
B_1^1 = \begin{bmatrix}
0.1 \\
1.1
\end{bmatrix}, \quad
B_2^1 = \begin{bmatrix}
0.1 \\
1.1
\end{bmatrix}.
$$

The system matrices for the second agent are:
$$
A_1^2 = \begin{bmatrix}
1 & 0.01 \\
0.5 & 1
\end{bmatrix}, \quad
A_2^2 = \begin{bmatrix}
1 & 0.01 \\
0.55 & 1
\end{bmatrix}, \quad
B_1^2 = \begin{bmatrix}
0.1 \\
0.1
\end{bmatrix}, \quad
B_2^2 = \begin{bmatrix}
0.1 \\
0.5
\end{bmatrix}.
$$

The system matrices for the third agent are:
$$
A_1^3 = \begin{bmatrix}
0.8 & 0.01 \\
0.1 & 1
\end{bmatrix}, \quad
A_2^3 = \begin{bmatrix}
0.75 & 0.01 \\
0.1 & 1
\end{bmatrix}, \quad
B_1^3 = \begin{bmatrix}
0.1 \\
1.1
\end{bmatrix}, \quad
B_2^3 = \begin{bmatrix}
0.1 \\
1.1
\end{bmatrix}.
$$

The system matrices for the fourth agent are:
$$
A_1^4 = \begin{bmatrix}
1 & 0.01 \\
0.5 & 1
\end{bmatrix}, \quad
A_2^4 = \begin{bmatrix}
1 & 0.01 \\
0.45 & 1
\end{bmatrix}, \quad
B_1^4 = \begin{bmatrix}
0.1 \\
0.1
\end{bmatrix}, \quad
B_2^4 = \begin{bmatrix}
0.1 \\
0.5
\end{bmatrix}.
$$

The dynamic event-triggering mechanism parameters for the four agents are as follows:
$$
\alpha_1 = \alpha_3 = 0.02, \quad \alpha_2 = \alpha_4 = 0.03,
$$
$$
\beta_1 = \beta_2 = \beta_3 = \beta_4 = 0.5,
$$
$$
\theta_1 = \theta_3 = 0.02, \quad \theta_2 = \theta_4 = 0.3.
$$

\begin{figure}[!ht]
    \centering
    \includegraphics[width=0.7\textwidth]{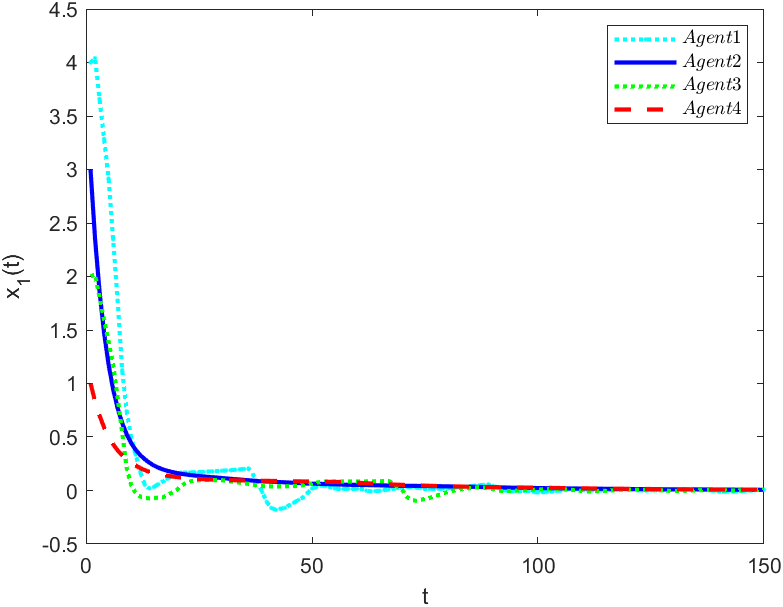}
    \caption{The state $x_1$ responses of the MAS.}
    \label{fig:4.3}
\end{figure}

\begin{figure}[!ht]
    \centering
    \includegraphics[width=0.7\textwidth]{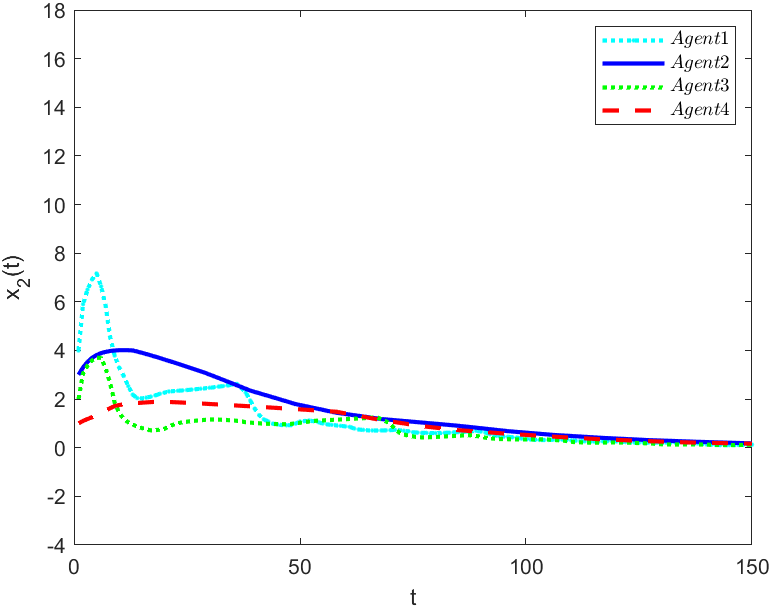}
    \caption{The state $x_2$ responses of the MAS.}
    \label{fig:4.4}
\end{figure}

Additionally, let $\kappa = 2$. According to Theorem 4.1, the controller gains can be obtained as follows:
$$
\left\{
\begin{aligned}
&\text{Agent 1:} & K_1^{1(2)} &= [-0.003042760, -0.016776025], & K_1^{1(1)} &= [0.010595568, 0.002936131], \\
& & K_2^{1(2)} &= [-0.003042756, -0.016776031], & K_2^{1(1)} &= [0.010595559, 0.002936133]; \\
&\text{Agent 2:} & K_1^{2(2)} &= [-0.548400402, -0.520870655], & K_1^{2(1)} &= [0.251879344, 0.003853859], \\
& & K_2^{2(2)} &= [-0.548400809, -0.520870819], & K_2^{2(1)} &= [0.251879100, 0.003853939]; \\
&\text{Agent 3:} & K_1^{3(2)} &= [-0.002068190, -0.014209980], & K_1^{3(1)} &= [0.008264478, 0.002771699], \\
& & K_2^{3(2)} &= [-0.002068182, -0.014209974], & K_2^{3(1)} &= [0.008264475, 0.002771695]; \\
&\text{Agent 4:} & K_1^{4(2)} &= [-0.588968241, -0.500429766], & K_1^{4(1)} &= [0.275143249, -0.018194835], \\
& & K_2^{4(2)} &= [-0.588967974, -0.500429594], & K_2^{4(1)} &= [0.275143351, -0.018194883].
\end{aligned}
\right.
$$

The initial states of each independent agent are:
$$
x_1(t_0) = \begin{bmatrix} 1 \\ 1 \end{bmatrix}, \quad
x_2(t_0) = \begin{bmatrix} 2 \\ 2 \end{bmatrix}, \quad
x_3(t_0) = \begin{bmatrix} 3 \\ 3 \end{bmatrix}, \quad
x_4(t_0) = \begin{bmatrix} 4 \\ 4 \end{bmatrix}.
$$

Based on the parameters given above and the obtained controller gains, the system response curves can be obtained.

\begin{figure}[!ht]
    \centering
    \includegraphics[width=0.7\textwidth]{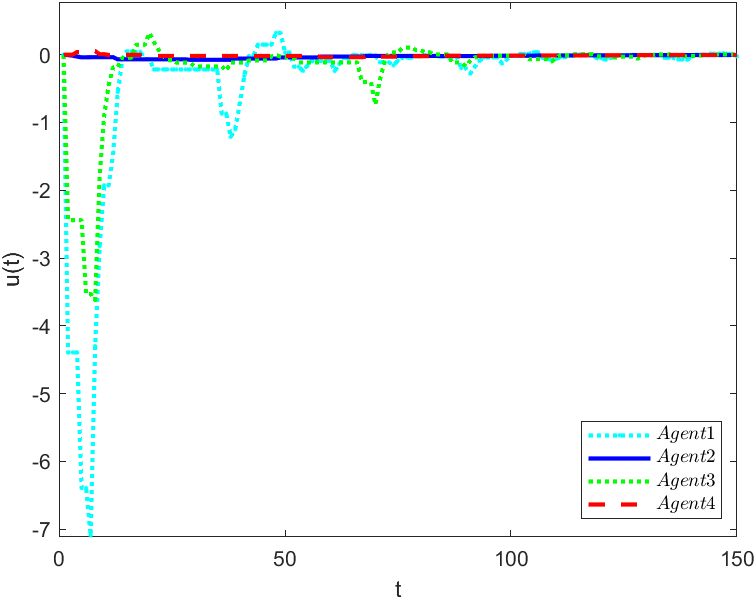}
    \caption{The control signal $u$ responses of the MAS.}
    \label{fig:4.5}
\end{figure}

The event-triggering situation of the multi-agent system is shown in Figure 4.2. In Figure 4.2, TRs denotes the triggered rates of the dynamic event-triggering mechanism (Triggered Rates, TRs), and its calculation method is as follows:
$$
TRs = \frac{\text{Total number of event triggering times}}{\text{Total number of sampling times}}.
$$

It can be observed that the TRs for the four agents are 0.46, 0.12, 0.39, and 0.11, respectively, indicating a significant reduction in communication burden.

The state response curves of the multi-agent system are illustrated in Figures 4.3 and 4.4, while the control signal response curves of the multi-agent control system are shown in Figure 4.5. From the simulation results, it can be observed that the controller design method effectively ensures consensus within the multi-agent system and stability of individual agents with fewer event triggers.

\section{Conclusions}

This study details the design of a memory-based dynamic event-triggered controller for heterogeneous MASs using IT2 T-S fuzzy models. The method addresses heterogeneity and uncertainties through IT2 T-S fuzzy modeling, introducing a novel dynamic threshold-based ETM that reduces communication overhead while maintaining performance. A Non-PDC strategy-based memory controller enhances robustness against delays and varying conditions. Controller design conditions were derived using a less conservative LMI transformation approach. Simulations confirmed the effectiveness in achieving consensus and stability with fewer event triggers, thereby enhancing MAS efficiency. Future research will explore adaptive mechanisms for real-time parameter changes and applications in large-scale MASs.

\printbibliography

\end{document}